# On the Fundamental Limits of Interweaved Cognitive Radios

G. Chung, and S. Vishwanath
Wireless Networking and Communication Group
University of Texas at Austin
Austin, TX 78712, USA
Email: {gchung,sriram}@ece.utexas.edu

C. S. Hwang
Communication Lab., SAIT
Samsung Electronics Co. Ltd.
Yongin, Korea
Email: cshwang@samsung.com

*Abstract*— This paper considers the problem of channel sensing in cognitive radios. The system model considered is a set of $N$ parallel (dis-similar) channels, where each channel at any given time is either available or occupied by a legitimate user. The cognitive radio is permitted to sense channels to determine each of their states as available or occupied. The end goal of this paper is to select the best $L$ channels to sense at any given time. Using a convex relaxation approach, this paper formulates and approximately solves this optimal selection problem. Finally, the solution obtained to the relaxed optimization problem is translated into a practical algorithm.
[1]

## I. INTRODUCTION

As the number and types of wireless (multimedia) applications increase, so do the stringent requirements they impose on the wireless medium. Thus, it is essential that we determine efficient means of utilizing limited spectral resources available to us. Currently, bandwidth resources are divided into frequency bands and allocated to different users exclusively in order to insure the quality of service (QoS) of multiple wireless systems, and the FCC's frequency allocation chart [1] shows that almost all frequency bands are currently divided and allocated to different groups for varying purposes. Also, according to recent surveys [2] and [3], most of this allocated radio frequency spectrum is vastly under-utilized. Cognitive radios are emerging as promising solutions to enable better utilization of spectrum especially in bands that are currently underutilized [4]. The classical example of a cognitive radio is one that employs "interweave" cognition [4]. These interweave cognitive radios are permitted to occupy a channel (frequency band) only when it is not occupied by a user licensed to use that band. If the presence of other radios can be sensed accurately and quickly, then such a policy can help ensure that cognitive users cause little to no interference to the licensed radios in the system. A majority of existing literature on cognitive radios focuses on such interweaved radios. For an analysis of other classes of cognitive radios, see [5], [6] and [7]. One of the main issues under study in the interweaved cognitive radio domain is the so-called "sensing problem", where we desire to determine, as accurately and efficiently as possible, if a given channel is occupied at any given time [8], [9], and [11]. For example, [8] describes a simple energy detection scheme for additive white Gaussian noise channel. In [9], the performance of energy detection schemes in a multipath environment is analyzed, and in [11], the impact of additional side information is considered in determining the performance of cognitive sensing. Overall, channel sensing is one of the better established fields of research on cognitive communication. In this paper, our goal is significantly different from that of channel-sensing literature. *Given a fairly accurate sensing algorithm, we desire to determine which channels should be sensed when*. In addition, we desire to perform a resource-allocation problem across multiple channels which may or may not be available to the cognitive radio. Overall, we ask the question "Given that there are multiple dissimilar channels available for you to sense, which channels should you sense and, if they are available, what rate/power should you assign to them?"

The dissimilarity between different channels arises from various factors. The properties of the propagation environment depend on frequency and thus can be significantly different from channel to channel. Some channels may suffer from "extraneous interference" from non-legitimate sources (such as in the industrial, scientific and medical (ISM) bands) that reduce the channel quality. Thus, just as any other multiband radio, the cognitive radio must allocate resources across different bands it uses while simultaneously determining which ones it is permitted to exploit. Note that, in isolation, the problem of channel selection for cognitive radios [12], [13] is well studied. Also, by itself, the resource allocation problem for muti-band radios is also well-understood [14]. However, bringing the two together is both important and challenging as they are tightly coupled in the context of interweaved cognitive radios. A simple explanation of this strong interdependence between sensing (channel selection) and resource allocation is as follows: Let us say that the system is such that "noisy" (poor) channels are less frequently used by licensed users than "clear" (good) channels. If the sensing mechanism were to choose to sense the infrequently-used channels, it will present the cognitive radio with available channels that are all "poor" resulting in a low rate. On the flip side, if the resource allocation mechanism were to assign high rates to the "good" channels, the sensing mechanism may find that they are not available for use and then again sustain a very poor rate. Thus, designing channel selection and allocation jointly is essential for cognitive radios. Note that this paper's focus is on the fundamental limits of joint selection and resource allocation in cognitive networks to provide a benchmark on performance. Thus, aspects such as sensing error, delay, device and network non-linearities etc. are not incorporated into the analysis.

[1] This work is supported by a grant from Samsung Advanced Institute of Technology.

Fig. 1. Channel Model

The rest of this is organized as follows. The next section details the system model and notations used in the paper. In Section III, we find the fundamental limit of the given system model. In Section IV, we propose an algorithm for joint channel selection and power allocation, and we conclude with Section IV.

## II. SYSTEM MODEL AND PROBLEM STATEMENT

The channel model is shown in Fig. 1. We consider $N$ parallel legitimate channels with equal bandwidth. In each time slot, a channel $n, 1 \leq n \leq N$, is occupied by a legitimate user with probability $q_n$. There are one cognitive transmitter and one cognitive receiver. The cognitive transmitter is allowed to transmit over channel $n$, if it is not occupied by any licensed user. In legitimate channel $n$, cognitive radio's channel is:

$$Y_n = X_n + Z_n$$

where $Z_n$ is additive Gaussian noise of variance $\sigma_n^2$. Note that this noise variance can be different from channel to channel, as it represents the fading state of that particular channel. Before the start of cognitive radio's transmission using the legitimate channels, the cognitive transmitter should know whether they are occupied by the licensed users or not. Thus, at the start of every time slot, the cognitive transmitter is allowed to sense a *subset* of channels, and is allowed to exploit those channels that are unoccupied; in this paper, we assume that the sensing is performed perfectly. Also, the cognitive transmitter is not allowed to transmit using the channel which is not sensed in order to guarantee the transmission of the licensed users. As $N$ is assumed to be large, it is impractical to allow the cognitive radio the ability to sense all of them at the start of every slot. Instead, we require it to cleverly choose a subset of bands on which to focus its efforts. The capacity of the cognitive radio depends on which channels to sense from $N$ parallel channels, and power allocation among the available parallel channels. Average total transmission power of cognitive transmitter is constrained to $P$.

First, define the $I_n(t)$ and $I_{E,n}(t)$ to be the indicator function for selected channel to be sensed and an indicator function for the unoccupied channel respectively, i.e.,

$$I_n(t) = \begin{cases} 0 & \text{if channel } n \text{ is not to be sensed} \\ 1 & \text{if channel } n \text{ is to be sensed} \end{cases} \quad (1)$$

and

$$I_{E,n}(t) = \begin{cases} 0 & \text{if channel } n \text{ is occupied} \\ 1 & \text{if channel } n \text{ is unoccupied} \end{cases}. \quad (2)$$

Denote the time average capacity of the cognitive radio with the selection of the sensing channel $I_n(t)$ and power allocation $P_n(t)$ in one time block as $\mathcal{C}(I_n(t), P_n(t))$. Then,

$$\mathcal{C}(I_n(t), P_n(t)) = \frac{1}{T} \sum_{n=1}^{N} \sum_{t=1}^{T} \frac{I_n(t) I_{E,n}(t)}{2} \log\left(1 + \frac{P_n(t)}{\sigma_n^2}\right), \quad (3)$$

where $T$ is the number of time slot in each time block.

In our model, we assume two constraints on the cognitive radio:

a. An average power constraint on the cognitive transmitter of $P$,

b. The number of channels that can be sensed by the cognitive radio at any given time is $L \leq N$.

Note that if the cognitive radio could sense all channels, $L = N$, this problem has a fairly trivial solution. At the start of each time slot, the cognitive radio would determine all available channels and *waterfill* its power over them [10].

If the number of channels that cognitive radio can sense is less than $N$, i.e. $L < N$, the resulting optimization problem can be stated as follows:

$$\max_{P_n(t), I_n(t)} \mathcal{C}(I_n(t), P_n(t)) \quad (4a)$$

such that

$$\frac{1}{T} \sum_{n=1}^{N} \sum_{t=1}^{T} I_n(t) I_{E,n}(t) P_n(t) \leq P, \quad (4b)$$

$$\sum_{n=1}^{N} I_n(t) \leq L, \quad (4c)$$

and

$$\begin{aligned} P_n(t) &\geq 0, \\ I_n(t) &\in \{0,1\}, \\ I_{E,n}(t) &\in \{0,1\}. \end{aligned} \quad (4d)$$

The optimization problem given by (4) determines the maximum empirical average rate achieved by the cognitive radio given constraints on the system. Note that it is an integer programming (IP) due to the constraints in (4d), and multi-dimensional due to its dependence on time $t$.

The next section studies the optimization problem given by (4) in an ergodic policy setting.

## III. Optimal Power Allocation and Selection of Sensing Channel

As a first step, we assume that our policy is ergodic and "static", i.e., that our sensing and power allocation policies are only functions of the channel statistics and do not evolve with time. This results in the following (simplified) optimization problem:

$$\max_{P_n, I_n} \sum_{n=1}^{N} \frac{I_n q_n}{2} \log\left(1 + \frac{P_n}{\sigma_n^2}\right) \quad (5a)$$

such that

$$\sum_{n=1}^{N} I_n q_n P_n \leq P, \quad (5b)$$

$$\sum_{n=1}^{N} I_n \leq L, \quad (5c)$$

and

$$P_n \geq 0, \quad I_n \in \{0, 1\}. \quad (5d)$$

Since $P_n = 0$ where $I_n = 0$, constraints (5b) can further be relaxed to

$$\sum_{n=1}^{N} q_n P_n \leq P. \quad (5e)$$

Denoting the optimal selection of channels to be sensed and power allocation for channel $n$ as $I_n^*$ and $P_n^*$ respectively, the optimum solution for (5) is given by the following theorem:

*Theorem 1:* The joint channel selection & rate allocation problem (characterized by the optimization problem in (5) is maximized when:

$$I_n^* = \arg\max_{I_n} \sum_{n=1}^{N} \frac{q_n I_n}{2} \left\lceil \log \frac{\lambda}{\sigma_n^2} \right\rceil^+$$

$$P_n^* = \left\lceil \lambda - \sigma_n^2 \right\rceil^+ I_n^*,$$

where

$$\sum_{n=1}^{N} \left\lceil \lambda - \sigma_n^2 \right\rceil^+ I_n^* q_n = P$$

$$\sum_{n=1}^{N} I_n^* = L,$$

and $\lceil w \rceil^+$ is maximum value of 0 and $w$.

**Proof:** Note that (5a) is a concave function over $P_n$ for a particular choice of $I_n$. The following Lagrangian describes the optimization of (5a) with respect to $P_n$ for a given $I_n$:

$$\mathcal{L} = \begin{array}{l} \sum_{n=1}^{N} \frac{I_n q_n}{2} \log\left(1 + \frac{P_n}{\sigma_n^2}\right) \\ -\lambda^{(1)}\left(\sum_{n=1}^{N} q_n P_n - P\right) + \sum_{n=1}^{N} \lambda_n^{(1)} P_n \end{array} \quad (6)$$

Taking the derivative of (6) and setting it to zero, we get:

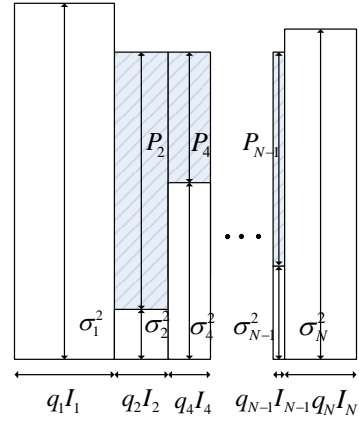

Fig. 2. power allocation with given $I_n$ ($I_1 = 1, I_2 = 1, I_3 = 0, ..., I_N = 1$)

$$\frac{\partial \mathcal{L}}{\partial P_n} = \frac{I_n q_n \log e}{2(P_n^* + \sigma_n^2)} - \lambda^{(1)} q_n + \lambda_n^{(1)} = 0. \quad (7)$$

$$P_n^* = \left\lceil \frac{I_n \log e}{2\lambda^{(1)}} - \sigma_n^2 \right\rceil^+ = \left\lceil \lambda - \sigma_n^2 \right\rceil^+ I_n, \quad (8)$$

where $\lambda = \frac{\log e}{2\lambda^{(1)}}$.

From (5e) we obtain,

$$\sum_{n=1}^{N} \left\lceil \lambda - \sigma_n^2 \right\rceil^+ I_n q_n = P. \quad (9)$$

Fig. 2. provides a graphical representation for the power allocation strategy in (8). Note that it is similar to the water-filling solution, with the main difference that the each channel has different width, $q_n I_n$. We refer to the policy in (9) as *modified* water-filling throughout this paper. Given that we understand the structure of the power allocation policy that optimizes (5a), we now desire to determine $I_n^*$. Note again that the optimization problem in (5) with respect to $I_n$ is an IP. It can be found by an exhaustive search, but computationally very hard to solve. Moreover, the power allocation strategy in 8, specifically, the water-level $\lambda$ is tightly coupled with the choice of $I_n$. In the next section, we present an algorithmic framework that approximates $I_n^*$ (and thus the water-level $\lambda$) using low-complexity iterative techniques.

## IV. Joint Selection and Power Control

A typical integer program is non-polynomial in complexity. Although multiple techniques exist for obtaining approximate solutions to such a program (such as branch and bound [15], relaxation), such techniques apply to any integer program and do not take the structure of the problem into consideration. Our focus is on developing an algorithm customized to this problem setting. We perform this in two steps, which we call "coarse" and "fine" optimization. The coarse optimization step determines a set of $L$ channels to be utilized by the cognitive radio. It gives us the lowest possible waterlevel, $\lambda_{min}$. The fine optimization step uses $\lambda_{min}$ which we obtained from coarse optimization to further optimize the choice of the $L$ channels. First, we describe the coarse optimization step:

*Coarse Optimization*: We iteratively find the channels to sense along with modified water-filling which incur the lowest water level. Let $\lambda_{min}$ denote the lowest water level, and $I_n^c$ and $P_n^c$ indicate the selection of the channel and power allocation which result in $\lambda_{min}$. Detailed procedures to find $\lambda_{min}$, $I_n^c$, and $P_n^c$ is described in the following four steps.

Step I: Start with $L$ initial channels. We can choose $L$ channels with the largest $q_n$ as initial channels, for example.

$$I_{n,0} = \begin{cases} 1 & \text{if } q_n \text{ is among } L \text{ largests} \\ 0 & \text{otherwise} \end{cases} \quad (10)$$

$$S_0 = \{n \in [1, N] | I_{n,0} = 1\} \quad (11)$$
$$j = 1 \quad (12)$$

Step II: Perform the modified water-filling with $I_{n,j-1}$, $j \geq 1$, such that

$$\sum_{n=1}^{N} \lceil \lambda_j - \sigma_n^2 \rceil^+ I_{n,j-1} q_n = P. \quad (13)$$

Step III: Calculate $q_n(\lambda_j - \sigma_n^2)$, and select the largest $L$ channels.

$$I_{n,j} = \begin{cases} 1 & \text{if } \begin{array}{l} q_n(\lambda_j - \sigma_n^2) > 0 \ \& \\ q_n(\lambda_j - \sigma_n^2) \text{ is among } L \text{ largests} \end{array} \\ 0 & \text{otherwise} \end{cases}$$
$$(14)$$

$$S_j = \{n \in [1, N] | I_{n,j} = 1\} \quad (15)$$

Step IV: If $S_j = S_{j-1}$, terminate the iteration, and set the power allocation and channel selection values.

$$\lambda_{min} = \lambda_n \quad (16)$$
$$I_n^c = I_{n,j} \quad (17)$$
$$P_n^c = \left(\lambda_j - \sigma_n^2\right) I_{n,j}. \quad (18)$$

Otherwise, $j = j + 1$ and repeat from step II.

The coarse optimization is performed for two reasons. One is that the performance of coarse optimization is very close to the optimum. This will be shown from the simulation result in the next section. Here, the optimality of the coarse optimization in one special case will be stated and proven.

*Lemma 1:* Define $S_c$ to be the set of the channels which are selected from coarse optimization;

$$S_c = \{n \in [1, N] | I_n^c = 1\}.$$

If the noise variances of all the channels which are not selected in the coarse optimization are greater than the lowest water level $\lambda_{min}$, i.e.,

$$\sigma_n^2 \geq \lambda_{min}, \quad \forall n \in [1, N], n \notin S_c$$

then the coarse optimization is optimal.

**Proof:** Define $S^*$ to be the set of channels from optimal selection.

$$S^* = \{n \in [1, N] | I_n^* = 1\}.$$

From definition,

$$\max_{P_n} \sum_{n \in S^*} \frac{q_n}{2} \log\left(1 + \frac{P_n}{\sigma_n^2}\right) \geq \max_{P_n} \sum_{n \in S_c} \frac{q_n}{2} \log\left(1 + \frac{P_n}{\sigma_n^2}\right). \quad (19)$$

Let's assume that there exist at least one legitimate channel with noise variance higher than the lowest water level which is included in the optimal channel selection.

$$\exists n \in S^*, \ \sigma_n^2 \geq \lambda_{min}.$$

Define $S' = S^* \cup S_c$, and allow the number of channel to sense to be $M = |S^* \cup S_c|$, which are strictly larger than $L$. Then,

$$\max_{P_n} \sum_{n \in S'} \frac{q_n}{2} \log\left(1 + \frac{P_n}{\sigma_n^2}\right) \geq \max_{P_n} \sum_{n \in S^*} \frac{q_n}{2} \log\left(1 + \frac{P_n}{\sigma_n^2}\right). \quad (20)$$

Note that $S_c \subseteq S'$, and $\sigma_n^2 \geq \lambda_{min}$ for all $n \notin S_c$. Modified water-filling of $M$ channels in $S'$ will lead to $P_n=0$ for all $n \notin S_c$. Thus,

$$\max_{P_n} \sum_{n \in S'} \frac{q_n}{2} \log\left(1 + \frac{P_n}{\sigma_n^2}\right) = \max_{P_n} \sum_{n \in S_c} \frac{q_n}{2} \log\left(1 + \frac{P_n}{\sigma_n^2}\right).$$

Combine the above result with (20), we obtain

$$\max_{P_n} \sum_{n \in S_c} \frac{q_n}{2} \log\left(1 + \frac{P_n}{\sigma_n^2}\right) \geq \max_{P_n} \sum_{n \in S^*} \frac{q_n}{2} \log\left(1 + \frac{P_n}{\sigma_n^2}\right). \quad (21)$$

Above result contradict (19), unless $S_c$ is the optimal. This concludes the proof.

The other reason for performing the coarse optimization is that it provides the essential information, $\lambda_m in$, which is necessary for further fine optimization. Upon the following assumption, fine optimization is optimal.

*Conjecture 1:* If the noise variance $\sigma_n^2$ is greater than the water level in the coarse optimization ($\sigma_n^2 > \lambda_{min}$), then the channel $n$ is not likely to be sensed in the optimal strategy, or even if it is included it will not increase the capacity much;

**Intuition:** The following gives the intuition for the above conjecture. Define $S^+$ to be the set of channels with noise variance greater then or equal to the $\lambda_{min}$ and $S^-$ to be the set of channels with noise variance less then $\lambda_{min}$ but not included in $S_c$;

$$S^+ = \{n \in [1, N] | \sigma_n^2 > \lambda_{min}\},$$
$$S^- = \{n \in [1, N] | \sigma_n^2 \leq \lambda_{min}, n \notin S_c\}.$$

We have the set of channel $S_c$ which incur the lowest waterlevel. Lemma(1) shows that average capacity cannot increase by exchanging elements in $S^+$ with elements in

$S_c$. Thus, for elements in $S^+$ to be included in $S^*$, optimal channel selection, elements in $S^-$ should be included also. By exchanging elements in $S^-$ with elements in $S_c$ the waterlevel rises up. Elements in $S^+$ can only be in optimal selection if exchanging $S^+$ with elements in $S_c$ lower the waterlevel which increased due to the inclusion of channels in $S^-$ effectively. Intuition is that channels in $S_c$ are the channels which can lower the waterlevel effectively already. Thus, effect of lowering the waterlevel with channels in $S^+$ will not affect much in increasing the average capacity. The validity of this conjecture is shown from the numerical analysis.

*Fine Optimization*: From lemma(1), if the number of channels that is selected to sense from the coarse optimization is less than $L$, it is optimal, and no further optimization is needed. Otherwise, further optimization will be required. From Conjecture(1), we reconstruct the problem, so that we can optimize the selection of the channel over the channels with noise variance smaller than or equal to $lambda_min$ only. we rearrange the useful channels by indexing from 1 to $M$, where $M$ is the number of channels that has noise variance smaller than $\lambda_{min}$;

$$M = |S_c \cup S^-| \quad (22)$$
$$\sigma_n^2 - \lambda_{min}/le0 \quad \forall n \in [1, M]. \quad (23)$$

Then, the optimization problem can be rewritten as follows;

$$\max_{\lambda, I_n} \mathcal{C}(\lambda, I_n) = \max_{\lambda, I_n} \sum_{n=1}^{M} \frac{q_n}{2} \log\left(1 + \frac{\lceil \lambda - \sigma_n^2 \rceil^+ I_n}{\sigma_n^2}\right) \quad (24)$$

$$= \max_{\lambda, I_n} \sum_{n=1}^{M} \frac{q_n I_n}{2} \log\left(1 + \frac{\lceil \lambda - \sigma_n^2 \rceil^+}{\sigma_n^2}\right) \quad (25)$$

$$= \max_{\lambda, I_n} \sum_{n=1}^{M} \frac{q_n I_n}{2} \left[\log \frac{\lambda}{\sigma_n^2}\right]^+, \quad (26)$$

$$\stackrel{(a)}{=} \max_{\lambda, I_n} \sum_{n=1}^{M} \frac{q_n I_n}{2} \log \frac{\lambda}{\sigma_n^2}, \quad (27)$$

such that

$$\sum_{n=1}^{M} \lceil \lambda - \sigma_n^2 \rceil^+ I_n q_n \stackrel{(b)}{=} (\lambda - \sigma_n^2) I_n q_n = P, \quad (28)$$

$$\sum_{n=1}^{M} I_n \leq L, \quad (29)$$

$$\lambda \geq \lambda_{min} \quad (30)$$

$$I_n \in \{0, 1\}, \quad (31)$$

where $(a)$ and $(b)$ result from constraining $\lambda \geq \lambda_{min}$. Then the optimal channel selection and power allocation can be determined by using the following theorem.

*Theorem 2:*

$$\lambda = \frac{\sum_{n=1}^{M} q_n I_n \sigma_n^2 + P}{\sum_{n=1}^{M} q_n I_n}$$

$$I_n = 0 \quad \text{if} \quad \lambda > \sigma_n^2 e^{1 - \frac{\sigma_n^2}{\lambda}}$$

**Proof:** Relax the constraint on $I_n$, such that the $I_n$ can take the value in the region $[0, 1]$. Construct the objective function $\mathcal{C}(\lambda, f(I_n))$ such that it is concave over the region of $I_n$ and $\lambda$, and $f(0) = 0$ and $f(1) = 1$. Consider the function $f(I_n) = I_n^k$, then $f(0) = 0$, $f(1) = 1$, and

$$\mathcal{C}(\lambda, f(I_n)) = \sum_{n=1}^{M} \frac{q_n I_n^k}{2} \log \frac{\lambda}{\sigma_n^2}. \quad (32)$$

Concavity of $\mathcal{C}(\lambda, f(I_n))$ can be found as follows;

$$\begin{bmatrix} \frac{\partial^2 \mathcal{C}(\lambda, f(I_n))}{\partial I_n^2} & \frac{\partial^2 \mathcal{C}(\lambda, f(I_n))}{\partial I_n \partial \lambda} \\ \frac{\partial^2 \mathcal{C}(\lambda, f(I_n))}{\partial \lambda \partial I_n} & \frac{\partial^2 \mathcal{C}(\lambda, f(I_n))}{\partial^2 \lambda} \end{bmatrix} \quad (33)$$

$$= \begin{bmatrix} q_n k (k-1) I_n^{k-2} \log \frac{\lambda}{\sigma_n^2} & q_n k I_n^{k-1} \frac{1}{\lambda} \log e \\ q_n k I_n^{k-1} \frac{1}{\lambda} \log e & -\sum_{j=1}^{M} q_j I_j^k \frac{1}{\lambda^2} \log e \end{bmatrix}. \quad (34)$$

Since the matrix is symmetric, if the determinant, $(1, 1)$ and $(2, 2)$ components of the matrix take the negative values, the matrix is negative semi-definite. Thus,

$$k(k-1) \leq 0 \quad (35)$$

$$k(k-1) \log \frac{\lambda}{\sigma_n^2} + k^2 \log e \leq 0 \quad (36)$$

are the condition for $\mathcal{C}(\lambda, f(I_n))$ to be a concave function. We can find $k$ such that the condition can be satisfied;

$$k = \min_{\sigma_n^2} \left( \frac{\log \frac{\lambda_{min}}{\sigma_n^2}}{\log \frac{\lambda_{min}}{\sigma_n^2} + \log e} \right). \quad (37)$$

Now that we verified the concavity of the objective function, we can construct the according Lagrangian multiplier:

$$\mathcal{L} = \begin{array}{l} \sum_{n=1}^{M} \frac{q_n I_n^k}{2} \log\left(\frac{\lambda}{\sigma_n^2}\right) \\ -\mu_0 \left( \sum_{n=1}^{M} q_n I_i^k (\lambda - \sigma_n^2) - P \right) - \mu_1 \left( \sum_{n=1}^{M} I_i^k - L \right) \\ + \sum_{n=1}^{N} \mu_{2,i} I_i - \sum_{n=1}^{N} \mu_{3,i} (I_i - 1) + \mu_4 (\lambda - \lambda_{min}). \end{array} \quad (38)$$

Solving the optimization,

$$\begin{aligned} \frac{\partial \mathcal{L}}{\partial I_n} &= \begin{array}{l} k q_n I_n^{k-1} \log \frac{\lambda}{\sigma_n^2} - \mu_0 q_n k I_n^{k-1} (\lambda - \sigma_n^2) - \mu_1 k I_n^{k-1} \\ + \mu_{2,i} - \mu_{3,i} \end{array} \\ &= 0 \end{aligned} \quad (39)$$

$$\frac{\partial \mathcal{L}}{\partial \lambda} = \sum_{n=1}^{M} q_n I_n^k \frac{\log e}{\lambda} - \mu_0 \sum_{n=1}^{M} q_n I_n^k + \mu_4 = 0 \quad (40)$$

$$\mu_0 \left( \sum_{n=1}^{M} q_n I_n^k (\lambda - \sigma_n^2) - P \right) = 0, \quad (41)$$

$$\mu_1 \left( \sum_{n=1}^{M} I_n^k - L \right) = 0, \quad (42)$$

$$\mu_{2,i} I_n = 0, \quad (43)$$

$$\mu_{3,i}(I_n - 1) = 0, \qquad (44)$$

$$\mu_4(\lambda - \lambda_{min}) = 0, \qquad (45)$$

where $\mu_0$, $\mu_1$, $\mu_{2,i}$, $\mu_{3,i}$, and $\mu_4$ are non-negative values. From the condition (40),

$$\lambda = \frac{\sum_{n=1}^{M} q_n I_n^k \log e}{\mu_0 \sum_{n=1}^{M} q_n I_n^k - \mu_4}. \qquad (46)$$

Note that $\sum_{n=1}^{M} q_n I_n^k \log e$, $\sum_{n=1}^{M} q_n I_n^k$, and $\mu_4$ are the non-negative value. Thus, $\mu_0$ should be positive number in order for $\lambda$ to be positive. Then, from the condition (41),

$$\sum_{n=1}^{M} q_n I_n^k (\lambda - \sigma_n^2) = P \qquad (47)$$

Thus,

$$\lambda = \frac{\sum_{n=1}^{M} q_n I_n^k \sigma_n^2 + P}{\sum_{n=1}^{M} q_n I_n^k}, \qquad (48)$$

and from rearranging the equation (46),

$$\mu_0 = \frac{\log e}{\lambda} + \frac{\mu_4}{\sum_{n=1}^{M} q_n I_n^k}. \qquad (49)$$

From the condition (39), we find that

$$I_n = \left( \frac{\left( q_n \log \frac{\lambda}{\sigma_n^2} - \mu_0 q_n (\lambda - \sigma_n^2) - \mu_1 \right) k}{\mu_{3,i} - \mu_{2,i}} \right)^{\frac{1}{1-k}} \qquad (50)$$

If $I_n$ is not either 0 or 1, from conditions (43) and (44), $\mu_{2,i}$ and $\mu_{3,i}$ becomes 0, which will result in making $I_n$ infinite number. Thus, $I_n$ takes either 0,1 value, which gives the desirable solution, such that the optimization of the relaxed condition coincides with the condition for the original problem, and

$$\lambda = \frac{\sum_{n=1}^{M} q_n I_n \sigma_n^2 + P}{\sum_{n=1}^{M} q_n I_n}, \qquad (51)$$

We set $\mu_4$ to be zero, then from (49) and (50), we obtain

$$I_n = \frac{\left( q_n \log \frac{\lambda}{\sigma_n^2 e^{1-\frac{\sigma_n^2}{\lambda}}} - \mu_1 \right) k}{\mu_{3,i} - \mu_{2,i}}^{\frac{1}{1-k}}. \qquad (52)$$

As a result, $I_n$ can be 1 only if

$$\lambda > \sigma_n^2 e^{1 - \frac{\sigma_n^2}{\lambda}}. \qquad (53)$$

This conclude the proof. With the Theorem(2), we can design iterative algorithm to find the optimal selection of channels to sensed iteratively.

Step I: Set the channels from coarse optimization to be the initial channels.

$$I_{n,0} = \begin{cases} 1 & \text{if } n \in S_c \\ 0 & \text{otherwise} \end{cases} \qquad (54)$$

$$S_0 = \{n \in [1,N] | I_{n,0} = 1\} \qquad (55)$$

$$j = 1 \qquad (56)$$

Step II: Calculate the waterlevel $\lambda_j$ from (51)

$$\lambda_j = \frac{\sum_{n=1}^{M} q_n I_{n,j-1} \sigma_n^2 + P}{\sum_{n=1}^{M} q_n I_{n,j-1}}. \qquad (57)$$

Step III: Calculate $\lambda_j - \sigma_n^2 e^{1 - \frac{\sigma_n^2}{\lambda_j}}$, and select the largest $L$ channels.

$$I_{n,j} = \begin{cases} 1 & \text{if } \lambda_j - \sigma_n^2 e^{1-\frac{\sigma_n^2}{\lambda_j}} \text{ is among } L \text{ largests} \\ 0 & \text{otherwise} \end{cases} \qquad (58)$$

$$S_j = \{n \in [1,N] | I_{n,j} = 1\} \qquad (59)$$

Step IV: If $S_j = S_{j-1}$, terminate the iteration, and set the power allocation and channel selection values. $\lambda^f$, $I_n^f$, and $P_n^f$ are the waterlevel, channel selection, and power allocation from fine optimization, then

$$\lambda^f = \lambda_n \qquad (60)$$
$$I_n^f = I_{n,j} \qquad (61)$$
$$P_n^f = (\lambda_j - \sigma_n^2) I_{n,j}. \qquad (62)$$

Otherwise, $j = j + 1$ and repeat from step II. Following from theorem(**??**) this algorithm gives the optimal selection of the channels to be sensed and power allocation, with the assumption that the channels with noise variance greater than $\lambda_{min}$ do not affect the optimization.

## V. NUMERICAL ANALYSIS

In this section, we present numerical example of capacities for coarse and fine optimization along with optimal solution. In this example, dissimilarity between channels is implemented by adapting the multi-path fading, which will incur frequency selective channel. Also, occupation of the legitimate channel is modeled by having $q_n$ to be uniform in $[0,1]$ and identically distributed. Sixteen legitimate channels are considered, $N = 16$, and cognitive radio is allowed to select and sense eight channels from all the legitimate channels, $L = 8$. Fig. 3. compares the capacities of sub-optimal algorithms with the optimal one, where the performance of optimal channel selection is obtained from the exhaustive search. The graph shows that performance of the fine optimization meet with that of optimal one. Thus, it can be stated that the Conjecture (1) in fine optimization is valid. Coarse optimization also performs optimally in the low SNR region. In the low SNR region, it is likely that $\sigma_n^2 \geq \lambda_{min}, \ \forall n \in [1,N], n \notin S_c$, because there are not much power to waterfill. From the Lemma (1), coarse optimization is optimal in such case. It is also worthwhile to note that coarse optimization perform as well as the optimal one.

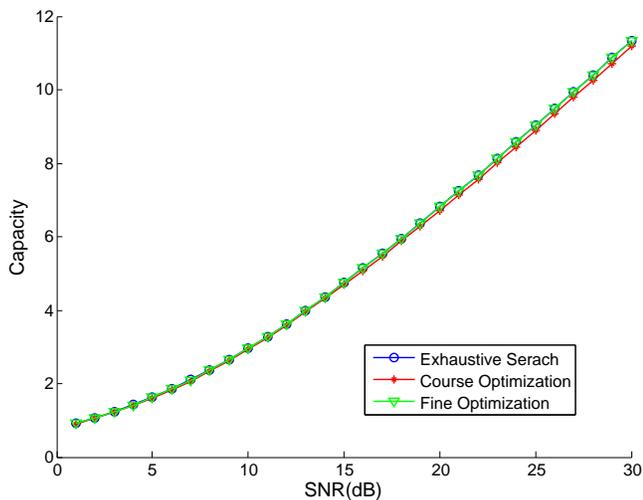

Fig. 3. Perforamance Analysis

## VI. CONCLUSION

In this paper, fundamental limits of interweaved cognitive radio has been verified. In the case that there are large number of legitimate channels and only limited number of them can be sensed, the capacity has been analyzed. However, it requires exhaustive search over the combination of all the legitimate channels, which is not practical in terms of complexity. Thus, two steps of sub-optimal solutions have been developed. Coarse optimization is developed, and verified to be optimal in the low SNR cases, and further optimization is performed to ensure the performance in the high SNR region also.

## ACKNOWLEDGMENT

The authors would like to thank Illsoo Sohn for useful discussions and comments.